\documentclass{article}
\usepackage{iclr2026_conference,times}

\usepackage{tabularx}
\usepackage{hyperref}
\usepackage{url}
\usepackage{booktabs}
\usepackage{enumitem}
\usepackage{cleveref}
\crefformat{section}{\S#2#1#3}
\crefformat{subsection}{\S#2#1#3}
\crefformat{subsubsection}{\S#2#1#3}
\crefformat{appendix}{Appx. #2#1#3}
\crefformat{table}{Tab.~#2#1#3}

\title{A Comparative Evaluation of AI Agent Security Guardrails}

\author{
\centerline{Qi Li \quad  Jiu Li \quad  Pingtao Wei \quad  Jianjun Xu \quad  Xueyi Wei} \\
\centerline{Jiwei Shi \quad  Xuan Zhang \quad  Yanhui Yang \quad  Xiaodong Hui \quad  Peng Xu \quad  Lingquan Zhou}
\\[0.5em]
\centerline{Beijing Caizhi Tech, Beijing, China}
\\[0.5em]
\centerline{liqi@czkj1010.com}
}

\setcitestyle{numbers}
\iclrfinalcopy

\begin{document}

\maketitle

\begin{abstract}
This report presents a comparative evaluation of DKnownAI Guard in AI agent security scenarios, benchmarked against three competing products: AWS Bedrock Guardrails, Azure Content Safety, and Lakera Guard. Using human annotation as the ground truth, we assess each guardrail's ability to detect two categories of risks: threats to the agent itself (e.g., instruction override, indirect injection, tool abuse) and requests intended to elicit harmful content (e.g., hate speech, pornography, violence). Evaluation results demonstrate that DKnownAI Guard achieves the highest recall rate at 96.5\% and ranks first in true negative rate (TNR) at 90.4\%, delivering the best overall performance among all evaluated guardrails.
\end{abstract}

\section{Evaluation Background}
\label{sec:background}

In our previous work \cite{deepknown2025}, we conducted an initial evaluation based on the S-Eval benchmark and our proprietary DeepKnown-High-Risk dataset, validating DKnownAI Guard's detection capabilities in general security scenarios.

The datasets used in that evaluation primarily covered traditional text content safety and did not adequately address the diverse attack scenarios that AI agents face in real-world deployments. As AI agents continue to evolve and gain widespread adoption, the security threats targeting them are accelerating in both scope and sophistication. The OpenClaw case serves as a compelling case study: OpenClaw is a widely-used AI agent application capable of directly controlling user computers through natural language, with high system privileges including file system read/write, environment variable management, API invocation, and plugin installation. Security researchers have disclosed multiple critical vulnerabilities in OpenClaw: attackers can execute prompt injection through malicious web pages to steal user credentials; manipulate the agent into deleting important data; compromise plugins and skill packages to exfiltrate API keys and deploy trojans; and the software itself contains multiple high-severity security vulnerabilities with notably inadequate default security configurations. These real-world cases demonstrate that the AI agent attack surface has expanded from traditional text content safety to multi-dimensional threats including instruction override, indirect injection, tool abuse, and plugin poisoning, with increasingly covert and complex attack techniques.

Driven by this trend, it is necessary to conduct more intensive evaluations across broader attack scenarios. This evaluation introduces multiple adversarial security datasets (see \cref{sec:methodology}), with emphasis on agent-specific attack scenarios including instruction override, indirect injection, role hijacking, chain-of-thought poisoning, and tool abuse. The attack intensity and deception level of these datasets significantly exceed those used in the previous evaluation, aiming to more comprehensively reflect the security challenges currently facing AI agents.

\section{Product Capabilities and Evaluation Objectives}
\label{sec:capabilities}

\subsection{DKnownAI Guard Core Capabilities}
\label{sec:core_capabilities}

DKnownAI Guard (\url{https://dknownai.com/}) provides comprehensive security protection for AI agent scenarios, covering two major categories of security capabilities.

\subsubsection{Agent Threat Detection}
Detects malicious inputs that attempt to control, exploit, or compromise the agent itself, preventing the agent from being weaponized as an ``attacker's tool.''

\begin{table}[h]
\centering
\caption{DKnownAI Guard Core Detection Capabilities}
\renewcommand{\arraystretch}{1.3}
\begin{tabularx}{\linewidth}{lX}
\toprule
\textbf{Capability} & \textbf{Description} \\
\midrule
Instruction Override Detection & Identifies direct or indirect attacks that override agent system instructions, including instruction replacement, delimiter attacks, and role hijacking \\
Privacy Data Leakage Prevention & Prevents attackers from inducing the agent to expose sensitive information such as passwords, API keys, and cryptographic keys \\
Malicious Behavior Manipulation Detection & Recognizes attacks that manipulate agent decision logic, including plan hijacking, chain-of-thought poisoning, and logical traps \\
Indirect Injection Detection & Detects malicious instructions embedded through contaminated external resources (web pages, documents, emails) accessed by the agent, including plugin and skill package poisoning \\
Tool Abuse Prevention & Prevents attackers from inducing the agent to invoke dangerous tools or APIs to execute destructive operations \\
\bottomrule
\end{tabularx}
\label{tab:core_capabilities}
\end{table}

\subsubsection{Harmful Content Detection}
Detects malicious requests intended to elicit inappropriate content from the agent, including hate speech, pornography, and violence, serving as a supplementary security capability.

\subsection{Product Advantages}

\begin{table}[h]
\centering
\caption{DKnownAI Guard Advantages}
\renewcommand{\arraystretch}{1.3}
\begin{tabularx}{\linewidth}{lX}
\toprule
\textbf{Dimension} & \textbf{Description} \\
\midrule
Dual-Channel Risk Classification & Independent detection channels that distinguish between ``agent threat'' and ``harmful content'' risks, enabling independent policy configuration for each risk type \\
Agent-Specific Detection & Dedicated detection engine optimized for agent-specific attack patterns including instruction override, privacy leakage, and behavior manipulation \\
Scenario-Based Configuration & Flexible security policy configuration by risk type, adaptable to diverse business scenario requirements \\
\bottomrule
\end{tabularx}
\label{tab:advantages}
\end{table}

\subsection{Evaluation Objectives}
\label{sec:objectives}

To validate DKnownAI Guard's practical protection effectiveness, this evaluation selects three competing products---AWS Bedrock Guardrails, Azure Content Safety, and Lakera Guard---for comparative testing. We assess each guardrail's detection capability for both agent threat security (instruction override, privacy data leakage, malicious behavior manipulation, indirect injection, tool abuse) and harmful content elicitation (hate speech, pornography, violence). Evaluation results are unified into a \textbf{BLOCKED / ALLOWED} binary classification, with human annotations serving as the ground truth for accuracy comparison.

\section{Evaluated Products}
\label{sec:products}

\begin{table}[h]
\centering
\caption{Evaluated Security Guardrail Products}
\renewcommand{\arraystretch}{1.3}
\begin{tabularx}{\linewidth}{llX}
\toprule
\textbf{Vendor} & \textbf{Product} & \textbf{Positioning} \\
\midrule
AWS & Bedrock Guardrails & LLM safety guardrail within the AWS ecosystem, supporting content filtering and contextual groundedness detection \\
Microsoft Azure & Content Safety & Part of Azure AI services, providing multi-modal (text/image) content safety moderation \\
Lakera & Lakera Guard & AI security startup specializing in prompt injection detection \\
DKnownAI & DKnownAI Guard & Agent security solution deeply optimized for AI agent scenarios, supporting dual-channel risk classification and scenario-based configuration \\
\bottomrule
\end{tabularx}
\label{tab:products}
\end{table}

\section{Evaluation Methodology}
\label{sec:methodology}

\subsection{Dataset Design}
\label{sec:datasets}

We randomly sampled \textbf{1,018} test entries from the following 8 public security datasets:

\begin{table}[h]
\centering
\caption{Evaluation Datasets}
\renewcommand{\arraystretch}{1.2}
\begin{footnotesize}
\begin{tabularx}{\linewidth}{lXl}
\toprule
\textbf{Dataset} & \textbf{Description} & \textbf{Attack Scenarios Covered} \\
\midrule
ALERT~\cite{alert2024} & Adversarial LLM prompt dataset & Agent threats \& harmful content \\
Salad-Data~\cite{salad2024} & Hierarchical safety benchmark with attack-enhanced queries & Jailbreak, harmful content \\
Tensor-Trust~\cite{tensortrust2023} & Human-generated prompt injection attacks from an online game & Prompt extraction, prompt hijacking \\
PromptWall-Injection~\cite{promptwall2025} & Prompt injection attack dataset & Instruction override, indirect injection \\
CSSBench~\cite{cssbench2026} & Contextual jailbreak benchmark & Role hijacking, chain-of-thought poisoning \\
UltraSafety~\cite{ultrasafety2024} & Harmful instructions with jailbreak prompts from AdvBench and AutoDAN & Jailbreak, harmful content \\
ToxicQAFinal~\cite{toxicqa2024} & Toxic question answering dataset & Harmful content \\
Jailbreak-Prompt-Injection~\cite{necent2026} & Aggregated dataset from 30+ public safety sources & Jailbreak, prompt injection, harmful content \\
\bottomrule
\end{tabularx}
\end{footnotesize}
\label{tab:datasets}
\end{table}

All datasets were originally annotated as malicious or harmful inputs. During the evaluation process, we conducted human re-annotation on top of the original labels, independently assessing the actual threat level of each entry: some entries originally labeled as harmful were determined not to pose actual threats in real business scenarios. Entries re-annotated as ALLOWED were retained in the evaluation without exclusion.

\subsection{Evaluation Procedure}
\label{sec:procedure}

\begin{enumerate}[leftmargin=*, topsep=0.3pt, itemsep=-0.5pt]
    \item \textbf{Human Re-annotation}: For the 1,018 randomly sampled entries, we conducted item-by-item human review based on the original dataset annotations, re-labeling each as \textbf{BLOCKED} (harmful) or \textbf{ALLOWED} (benign). Of these, 852 were labeled BLOCKED and 166 were labeled ALLOWED.
    \item \textbf{API Invocation}: All entries were sent to each security guardrail to obtain detection results.
    \item \textbf{Result Normalization}: Raw responses from each guardrail were unified into a \textbf{BLOCKED / ALLOWED} binary classification, aligned with human annotations.
    \item \textbf{Comparative Assessment}: Each guardrail's classification results were compared against human annotations to calculate recall rate and true negative rate.
\end{enumerate}

\section{Experimental Results}
\label{sec:results}

\subsection{Comprehensive Comparison}
\label{sec:comparison}

Using human annotations as the ground truth, the recall rate and true negative rate of each guardrail are shown in \cref{tab:results}.

\begin{table}[h]
\centering
\caption{Comprehensive Comparison Results (Human Annotation as Ground Truth)}
\renewcommand{\arraystretch}{1.4}
\begin{tabular}{lcccc}
\toprule
\textbf{Metric} & \textbf{AWS} & \textbf{Azure} & \textbf{DKnownAI} & \textbf{Lakera} \\
\midrule
Recall (BLOCKED, 852) & 743 (87.2\%) & 715 (83.9\%) & \textbf{822 (96.5\%)} & 812 (95.3\%) \\
True Negative Rate (ALLOWED, 166) & 149 (89.8\%) & 142 (85.5\%) & \textbf{150 (90.4\%)} & 145 (87.3\%) \\
\bottomrule
\end{tabular}
\label{tab:results}
\end{table}

DKnownAI Guard achieves the best overall performance, with a recall rate of 96.5\% and a true negative rate of 90.4\%. Lakera Guard demonstrates strong recall at 95.3\%, ranking second. AWS Guardrails achieves a true negative rate of 89.8\%, ranking second in TNR. Azure Content Safety shows relatively lower performance on both metrics.

\subsection{Evaluation Difficulty and True Negative Rate Analysis}
\label{sec:difficulty}

The true negative rate for some guardrails in this evaluation is relatively low, which falls within the expected range. The datasets introduced in this evaluation significantly exceed conventional evaluations in both attack intensity and deception level (see \cref{sec:background}). The ALLOWED samples are boundary cases selected through systematic human review from predominantly harmful datasets, inherently carrying partial semantic features of harmful data with strong ambiguity. The false positive rate for such high-ambiguity samples is significantly higher than for ordinary benign data. Therefore, the lower true negative rate for some guardrails is a characteristic of the evaluation data distribution rather than a deficiency in the guardrails themselves. Under these conditions, DKnownAI Guard maintains a 90.4\% true negative rate, demonstrating its superior ability to distinguish highly deceptive boundary samples compared to other vendors.

\section{Conclusions}
\label{sec:conclusions}

DKnownAI Guard achieves the best overall performance in this evaluation, ranking first in both recall rate and true negative rate.

At the same time, the relatively low true negative rate of Azure Content Safety reflects a common deficiency in current security guardrails when dealing with high-ambiguity boundary data. Even the best-performing guardrail still misblocks approximately 10\% of benign data, which in real-world deployments may result in legitimate user requests being blocked. Improving the classification precision of security guardrails on high-ambiguity boundary samples---maintaining high detection capability while effectively controlling the false positive rate---remains a critical challenge for the AI agent security community.

\appendix
\section{BLOCKED/ALLOWED Mapping Logic}
\label{sec:appendix_mapping}

All four guardrails employ dual-channel detection. An input is mapped to \textbf{BLOCKED} if either channel triggers, and to \textbf{ALLOWED} only if neither channel triggers. Both channels may trigger simultaneously.

\textbf{AWS Bedrock Guardrails.}
\begin{itemize}[leftmargin=*, topsep=0.3pt]
    \item \textbf{BLOCKED}: Any content policy filter is triggered (\texttt{PROMPT\_ATTACK}, \texttt{HATE}, \texttt{VIOLENCE}, \texttt{MISCONDUCT}, \texttt{INSULTS}, or \texttt{SEXUAL}).
    \item \textbf{ALLOWED}: No filter is triggered.
\end{itemize}

\textbf{Azure Content Safety.}
\begin{itemize}[leftmargin=*, topsep=0.3pt]
    \item \textbf{BLOCKED}: The \texttt{shieldPrompt} endpoint detects an attack, or the \texttt{text:analyze} endpoint returns any harmful category with severity $> 0$.
    \item \textbf{ALLOWED}: Neither endpoint reports a detection.
\end{itemize}

\textbf{Lakera Guard.}
\begin{itemize}[leftmargin=*, topsep=0.3pt]
    \item \textbf{BLOCKED}: A \texttt{prompt\_attack} or \texttt{moderated\_content/*} detector returns a high-confidence result (\texttt{l1\_confident} or \texttt{l2\_very\_likely}).
    \item \textbf{ALLOWED}: All detections fall below the confidence threshold or no detection occurs.
\end{itemize}

\textbf{DKnownAI Guard.}
\begin{itemize}[leftmargin=*, topsep=0.3pt]
    \item \textbf{BLOCKED}: The detection status is \texttt{AGENT\_HACK}, \texttt{SYS\_FLAG}, or \texttt{CONTENT\_FLAG}.
    \item \textbf{ALLOWED}: The detection status is any other value.
\end{itemize}

\end{document}